\documentclass[useAMS,usenatbib]{mn2e}
\usepackage{graphicx,lscape,amssymb}
\usepackage{natbib}
\usepackage{epstopdf}
\usepackage{tabulary}
\usepackage{fancyhdr}
\usepackage{subfigure}
\usepackage{booktabs, dcolumn}
\usepackage{arydshln}

\newcommand{\aj}{AJ}
\newcommand{\apj}{ApJ}
\newcommand{\apjs}{ApJ}
\newcommand{\apjl}{ApJ}
\newcommand{\aap}{A\&A}

\newcommand{\aapr}{A\&A~Rev.}         
\newcommand{\pasp}{PASP}

\title[Variable white dwarfs in SDSS Stripe 82]{A search for variable white dwarfs in large area time domain surveys: a pilot study in SDSS Stripe 82}

\author[Gentile Fusillo et al.]{Nicola Pietro Gentile Fusillo $^1$, J. J. Hermes $^1$, Boris T. G\"ansicke $^1$ \\$^1$ Department of Physics, University of Warwick, Coventry, CV4 7AL, UK\\}

\begin{document}
\maketitle

\label{firstpage}

\begin{abstract}
We present a method to reliably select variable white dwarfs  from large area time domain surveys and apply this method in a pilot study to search for pulsating white dwarfs in the Sloan Digital Sky Survey Stripe\,82. From a sample 400 high-confidence white dwarf candidates, we identify 24 which show significant variability in their multi-epoch Stripe\,82 data. Using colours, we further selected a sample of pulsating white dwarf (ZZ Ceti) candidates  and obtained high cadence follow up for six targets. We confirm five of our candidates as cool ZZ Cetis, three of which are new discoveries. Among our 24 candidates we also identify: one eclipsing binary, two magnetic white dwarfs and one pulsating PG1159 star. Finally we discuss the possible causes for the variability detected in the remaining targets. Even with sparse multi-epoch data over the limited area of Stripe\,82, we demonstrate that our selection method can successfully identify  various types of variable white dwarfs and efficiently select high-confidence ZZ Ceti candidates.
\end{abstract}


\section{Introduction}
Many white dwarfs show some degree of variability in their apparent brightness. These brightness changes can be very different in nature: bright eruptions in interacting binaries (cataclysmic variables or classical novae; \citealt{warner95-1}),  eclipsing binaries (e.g \citealt{greenetal78-1, orosz+wade99-1, parsonsetal15-1}), rotationally variable magnetic white  dwarfs (\citealt{barstowetal95-2, brinkworthetal04-1, brinkworthetal13-1}; \citealt{lawrieetal13-1}), pulsating white dwarfs \citep{lasker+hesser71-1, bergeronetal95-1, mukadametal04-1, nittaetal09-1} and even systems in which the cause of variability is still unexplained \citep{maozetal15-1, holbergetal11-2}.
Each one of these classes offers a unique and different channel to explore white dwarf structure and evolution.
Pulsating white dwarfs, in particular, are extraordinary tools to probe their interior structure. 

Traditionally, the physical parameters of white dwarfs, including their effective temperature ($T_{\mathrm{eff}}$) and surface gravity ($\log g$) are determined from spectroscopic analysis \citep{bergeronetal92-2}. However, spectral information is restricted to the outermost layers of of the star. As a consequence, our understanding of white dwarfs is often, literally, superficial.

The existence of pulsating white dwarfs, however, provides a unique opportunity to probe the interior of these objects.
Asteroseismology can be used to  investigate 
the structure, composition and mass of both the core and envelope (\citealt{winget+kepler08-1}; \citealt{fontaine+brassard08-1}; \citealt{althausetal10-1}), internal rotation profiles \citep{charpinetetal09-1}, measure weak magnetic fields \citep{wingetetal91-1} and even search for planetary companions via pulse timing variations (\citealt{wingetetal03-1}; \citealt{mullallyetal08-1}).
Since most, if not all, white dwarfs evolve through a phase of pulsations as they cool, asteroseismological studies can shed light on the internal structure  of the global white dwarf population (e.g. \citealt{robinson79-1, fontaineetal85-1, fontaineetal13-1}; \citealt{romeroetal12-1}).

Current observational evidence suggests that all hydrogen-atmosphere (DA), while cooling through the temperature range $\simeq12,500 - 11,000$\,K, will undergo global, non-radial pulsations (\citealt{bergeronetal04-1}; \citealt{gianninasetal11-1}).
These pulsating white dwarfs are known as DAV or ZZ\,Cetis.
Aside from ZZ\,Cetis, there at least two more classes of pulsating white dwarfs:
hot pre-white dwarfs (PG1159 or DOV; \citealt{mcgrawetal79-1}) and helium atmosphere (DB) white dwarfs (V777 Her or DBV; \citealt{wingetetal82-1}). Recent studies suggested that variable hot carbon atmosphere (DQ) 	 white dwarfs may constitute a further class of pulsators (\citealt{fontaineetal08-1}; \citealt{montgomeryetal08-1}), however the true nature of their variability is still matter of debate \citep{lawrieetal13-1}.

The first ZZ Ceti and indeed the the first pulsating white dwarf, DA HL\,Tau\,76, was serendipitously discovered in 1965 by \citet{landolt68-1}.
Since then,  approximately 180 similar objects have been identified and today ZZ Cetis are  by far the largest and best studied class of pulsating white dwarfs. 
Their pulsation periods typically range from 100 to 1400 seconds 
(\citealt{clemens93-2}; \citealt{mukadametal04-1}) and can reach up to 1.7 hours for
extremely low-mass ($M_{\mathrm{WD}}< 0.25 \mathrm{M_{\odot}}	$ ) white dwarfs \citep{hermesetal13-1}.
Historically, unambiguous identification of pulsating white dwarfs  required several hours of 
continuous high cadence photometry (e.g. \citealt{mukadametal04-1, nittaetal09-1}), which is observationally expensive. Candidate selection has so far relied on colours and/or $T_\mathrm{eff}$ and $\log g$, estimated from model fits to spectra, with efficiencies ranging from 30\% to 80\% (e.g. \citealt{mukadametal04-1}).

In recent years, the opportunity to repeatedly survey large areas of the sky has rapidly advanced the field of time-domain astronomy. 
Time-domain exploration of the sky is at the forefront of modern astronomy with many wide-field surveys in operation or soon to come on-line (eg. CRTS, \citealt{catalina14-1}; PTF, \citealt{PTF09-1}; EVRYSCOPE, \citealt{evryscope15-1}; Pan-STARRS, \citealt{panstarss14-1}; Gaia, \citealt{Gaia_tran14-1}; LSST, \citealt{LSST11-1}).
In order to fully exploit these vast resources we will need to develop efficient selection algorithms, e.g. a robust method to 
identify pulsating white dwarf candidates is needed to optimize high-cadence photometric follow-up.

Here, we investigate the feasibility of using multi-epoch photometry from large-area surveys to reliably identify pulsating white dwarf candidates based on Stripe\,82 of the Sloan Digital Sky Survey (SDSS). 
Several successful studies have made use of Stripe\,82 multi-epoch observations to search for variable objects (eg. \citealt{sesaretal07-1}; \citealt{bramichetal08-1};  \citealt{beckeretal11-2}). However these studies mainly focused on identifying large-amplitude variable sources (e.g eclipsing binaries or flaring stars) and the potential of identifying low amplitude variability  (like that of pulsating white dwarfs) has not yet been explored.
Starting from a sample of 400 high-confidence  white dwarfs candidates from the catalogue of \citet{gentilefusilloetal15-1}, we recover, recalibrate and quality-control all available multi-epoch photometry to identify variable candidates.
Even though Stripe\,82 offers only low and irregular cadence over a relatively limited area, our study demonstrates promising results. In the near future a similar methodology, applied to superior time-domain surveys (e.g. Pan-STARSS, Gaia and LSST) will completely change the way we identify high-amplitude variable stars, including pulsating white dwarfs.  
  
\section{SDSS Stripe 82}
\label{sdss_st82}
SDSS has obtained $ugriz$ multi-band photometry, in the magnitude range $g \simeq15-22$ mag, for over a third of the sky, and spectroscopic follow up of over 4 million objects \citep{alametal15-1}.

\begin{figure}
\includegraphics[width=\columnwidth]{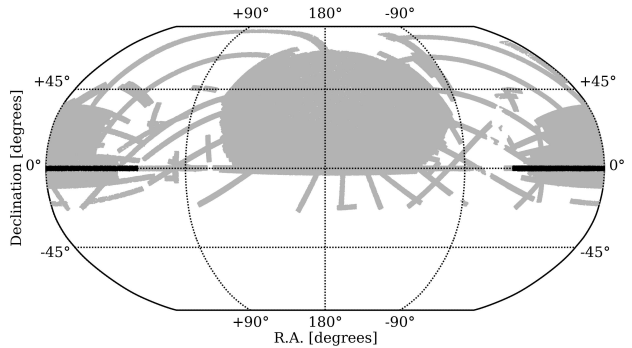}
\caption{\label{sky_stripe} Current photometric footprint of SDSS (data release 12, \citealt{alametal15-1}; $\simeq14,000\,\deg^2$). The $300\,\deg^2$ of Stripe\,82 are  shown in black.}
\end{figure}

A particular region of the SDSS footprint, Stripe\,82, has received multiple observations as part of different programs (most notably the SDSS-II Supernova survey,  \citealt{friemanetal08-1}; \citealt{sakoetal08-1}).
Stripe\,82 covers an area of $300\,\deg^2$  on the celestial equator spanning 8 hours in right ascension  ($-50^{\circ} <$ RA $< 59^{\circ}$) and 2.5\,degrees in declination ($-1.25^{\circ} <$ Dec $< 1.25^{\circ}$, see Fig.\,\ref{sky_stripe}). The stripe consists of two scan
regions which have been repeatedly imaged over approximately ten years with a total of 303 imaging runs. A specific imaging run may cover the entire length of the stripe or just a specific region, and several runs overlap in certain areas
\citep{abazajianetal09-1}. 

\section{Data selection and correction}\label{method_st82}
We retrieved photometry for 400 high-confidence white dwarf candidates from the catalogue of \citet{gentilefusilloetal15-1} which fall in Stripe\,82 and have \emph{probability of being a white dwarf} (P$_{\mathrm{WD}}$) $\geq0.75$.  We found that, on average, each white dwarf had been observed between 60 and 200 times. 
Only about one-quarter of the Stripe\,82 scans were obtained in photometric conditions; the rest were taken under variable clouds and often poorer than normal seeing. 
As a consequence, the default calibration of the Stripe\,82 photometry is insufficient to identify low-amplitude variables. In fact, all 400 white dwarf candidates (and their neighbouring objects) show significant low-amplitude variability (up to $\simeq 0.2$ mag) in their Stripe\,82  lightcurves.

In the standard procedure to correct for varying atmospheric conditions in time-series photometry, \textit{relative (or differential) photometry}, simultaneous observations of one or more neighbouring objects are used to calibrate the photometry of the science target. 
Changing atmospheric conditions affect neighbouring objects in the same way and can therefore be measured and corrected for. 
We followed this approach using the neighbouring stars of our white dwarf candidates to recalibrate the Stripe\,82 observations, and identify unreliable detections.
For each white dwarf candidate we retrieved Stripe\,82 photometry of all point sources with ``clean" photometry within a five arcminute radius, which were observed at the same time as the white dwarf.
We defined an ``individual nightly offset" as the difference between the $ugriz$ magnitudes of an object as detected on a given night and the median of all Stripe\,82 measurements of that object. Calculating the median of the ``individual nightly offsets" for all neighbours, then defined a ``median nightly offset" which we used to correct the Stripe\,82 magnitude of the white dwarf on that night.

However, several effects (such as irregular cloud coverage or intrinsic variability of a neighbour) can cause irregular variations from object to object, hence in certain nights the ``individual nightly offsets" are not consistent with each other and can therefore not be used to re-calibrate the field. 
We compare the ``individual nightly offsets" of all neighbours with the ``median nightly offset" and calculate reduced $\chi^2$ values. If a significant scatter is observed (${\chi^2}_{\mathrm{red}}>3$)
we conclude that the ``median night offset" cannot be considered reliable and discard the night. Furthermore we also consider unreliable those nights in which less than four neighbours were observed.


\begin{table*}
\caption{\label{phot_par}Stripe\,82 photometric parameters of the 26 variable white dwarf candidates and the three non-variable white dwarfs (below the dashed line)  selected as ``control" objects. Median $g$ magnitudes were calculated from all re-calibrated, reliable Stripe\,82 photometry. $g$ scatter were calculated via a Monte Carlo method, and represent the minimum expected amplitudes of variability required to cause the observed scatter in the Stripe\,82 photometry. Colours were calculated using median magnitudes.
Objects marked with $\dagger$ were later dropped as spurious candidates (see Sect.\,\ref{cont_st82} and Sect.\,\ref{err_pos}).} 
\begin{tabular}{lllcrrlc}
\hline
SDSS name & RA & Dec & median $g$ mag &  $u-g$& $g-r$ & $\chi^2$& $g$ scatter (mag) \\
\hline
\hline
SDSS\,J0028$-$0012 & 00 28 03.34 & $-$00 12 13.2 & 18.55 & 0.39 & $-$0.26 & 3.3 & 0.10 $\pm$ 0.02  \\
SDSS\,J0050$-$0001 & 00 50 13.52 & $-$00 01 30.3 & 18.64 & 0.13 & $-$0.31& 3.4 & 0.10 $\pm$ 0.01\\
SDSS\,J0050$-$0023 & 00 50 47.61 & $-$00 23 16.9 & 18.80 & 0.32 & $-$0.089 & 2.7 & 0.12 $\pm$ 0.02\\
SDSS\,J0102$-$0033 & 01 02 07.32 & $-$00 33 00.1 & 18.19 & 0.43 & $-$0.11 & 2.7 & 0.09 $\pm$ 0.02\\
SDSS\,J0106$-$0014 & 01 06 22.99 & $-$00 14 56.2 & 18.18 & 0.48 & $-$0.20 & 4.3 & 3.71 $\pm$ 0.39\\
SDSS\,J0121$-$0028 & 01 21 02.30 & $-$00 28 12.0 & 18.45 & 0.52 & $-$0.15 & 8.7 & 0.11 $\pm$ 0.02\\
SDSS\,J0134$-$0109 & 01 34 40.94 & $-$01 09 02.3 & 18.12 & 0.49 & $-$0.03 & 2.6 & 0.09 $\pm$ 0.02 \\
SDSS\,J0158$-$0000 & 01 58 01.11 & $-$00 00 00.2 & 18.62 & 0.36 & $-$0.22 & 5.1 & 0.16 $\pm$ 0.02 \\
SDSS\,J0209+0050 & 02 09 27.68 & +00 50 21.0 & 18.41 & $-$0.03 & $-$0.37 & 3.5 & 0.10 $\pm$ 0.02 \\
SDSS\,J0247+0003 & 02 47 46.29 & +00 03 31.6 & 16.26 & 0.24 & $-$0.38 & 3.4 & 0.07 $\pm$ 0.01 \\
SDSS\,J0318+0030 & 03 18 47.09 & +00 30 29.5& 17.86 & 0.42  & $-$0.13 & 2.7 &  0.08 $\pm$ 0.02\\
SDSS\,J0321$-$0050 & 03 21 43.49 & $-$00 50 25.6 & 18.89 & 0.27 & $-$0.26 & 2.4 & 0.10 $\pm$ 0.02 \\
SDSS\,J0326+0002 & 03 26 15.34 & +00 02 21.6 & 18.44 & 0.08 & $-$0.30 & 3.0 & 0.17 $\pm$ 0.02 \\
SDSS\,J0326+0018 & 03 26 19.44 & +00 18 17.5 & 17.41 &  0.39 & $-$0.20 & 7.8 & 0.13 $\pm$ 0.01 \\
SDSS\,J0342+0024$\dagger$ &  03 42 29.96 & +00 24 17.8 & 16.47 & 0.09 & $-$0.18 & 5.7 & 0.29 $\pm$ 0.02 \\

SDSS\,J0349$-$0059 & 03 49 17.40 & $-$00 59 19.2 & 17.65 & $-$0.39 & $-$0.34 & 2.7 &  0.07 $\pm$ 0.01\\
SDSS\,J2109+0111 & 21 09 33.63 & +01 11 10.6 & 18.96 & 0.12 &  $-$0.33 & 3.0 & 0.13 $\pm$ 0.02\\
SDSS\,J2156$-$0046 & 21 56 28.27 & $-$00 46 17.2 & 18.38 & 0.55 &  $-$0.06 & 3.8 & 0.07 $\pm$ 0.01\\
SDSS\,J2157+0037 & 21 57 13.51 & +00 37 14.8 & 17.41 & 0.51 & $-$0.12 & 9.3 & 0.11 $\pm$ 0.01\\
SDSS\,J2157$-$0044$\dagger$ & 21 57 11.87 & $-$00 44 34.9 & 18.63& $-$0.35 & $-$0.39 & 44.7 & 3.40 $\pm$ 0.25 \\
SDSS\,J2218$-$0000 & 22 18 28.59 & $-$00 00 12.2 & 18.09 & 0.15  & $-$0.20 & 2.2 & 0.07 $\pm$ 0.01 \\
SDSS\,J2220$-$0041 & 22 20 30.69 & $-$00 41 07.3 & 17.48 & 0.43 & 0.11 & 2.8 & 0.07 $\pm$ 0.01 \\
SDSS\,J2237$-$0101 & 22 37 26.85 & $-$01 01 10.8 & 18.88 & 0.46 & $-$0.11 & 4.8 &  0.12 $\pm$ 0.02 \\
SDSS\,J2318$-$0114 & 23 18 41.50 & $-$01 14 43.1 & 18.74 & 0.14 & $-$0.32 & 2.2 & 0.15 $\pm$ 0.02 \\
SDSS\,J2330+0100 & 23 30 40.50 & +01 00 47.6 & 17.52 & 0.66 & 0.25  & 3.2 & 0.13 $\pm$ 0.02  \\
SDSS\,J2333+0051 & 23 33 05.08 & +00 51 55.6 & 18.55 & $-$0.06 & $-$0.32 & 5.0 & 0.16 $\pm$ 0.02 \\\\
\hdashline\\
SDSS\,J0327+0012 & 03 27 27.52 & +00 12 52.5 & 17.83 & 0.44 & $-$0.17 & 0.8 & 0.04 $\pm$ 0.02 \\
SDSS\,J2245$-$0040 & 22 45 18.53 & $-$00 40 25.2 & 18.47 & 0.44 & $-$0.17 & 0.6 & 0.04 $\pm$ 0.02 \\
SDSS\,J2336$-$0051 & 23 36 47.00 & $-$00 51 14.6 & 18.32 & 0.48 & $-$0.20& 0.7 & 0.05 $\pm$ 0.02\\
\hline
\hline
\end{tabular}
\end{table*}

\begin{table*}
\caption{\label{lit_par} Additional parameters of the 26 white dwarf candidates initially identified as variable sources and the three non-variable white dwarfs (below the dashed line) selected as ``control" objects. Objects marked with $\dagger$ were later dropped as candidates. $T_{\mathrm{eff}}$ and $\log g$ are calculated using 1D atmospheric models. The initial class column shows the spectral classification  we assigned to the object based on the available SDSS spectrum. ``ZZ\,$_{\mathrm{cand}}$" are ZZ Ceti candidates (Sect.\,\ref{zz_c}). The ``LT obs" column shows 
the results of our LT time series follow up.    Object which were not observed to vary (in our LT follow up or by \citealt{mukadametal04-1}) are flagged as "NOV" followed by the non variability limit set by the observations.} 
\begin{tabular}{l D{?}{\pm}{7.7} D{?}{\pm}{7.7} lcc}
\hline
Object &  \multicolumn{1}{l}{$T_{\mathrm{eff}}$ $^a \mathrm{[K]}$} & \multicolumn{1}{c}{$\log g$ $^a$ [cgs]}& initial class & LT obs. &remarks\\
\hline
\hline \\ [-1.5ex]
SDSS\,J0028$-$0012 &  14,590?480 & 7.99?0.07  & DA & -- & --\\ 
SDSS\,J0050$-$0001 &  21,120 ? 570 & 7.59 ? 0.08  & DA & -- & --\\ 
SDSS\,J0050$-$0023 & 11,170 ? 90 & 8.84 ? 0.05 & ZZ\,$_{\mathrm{cand}}$ & -- & NOV0.6\,\%$^b$\\ 
SDSS\,J0102$-$0033 & 11,110 ? 170 & 8.37 ? 0.09 & ZZ\,$_{\mathrm{cand}}$ & ZZ Ceti & ZZ Ceti$^b$\\ 
SDSS\,J0106$-$0014 & 14,240 ? 300 & 7.49 ? 0.06 & ZZ\,$_{\mathrm{cand}}$ & -- & eclipsing binary$^c$\\ 
SDSS\,J0121$-$0028 & 10,447 ? 30^d & 8.42 ? 0.03^d & ZZ\,$_{\mathrm{cand}}$ & NOV1.4\,\% & --\\
SDSS\,J0134$-$0109 & 10,490 ? 80 & 8.04 ? 0.07 & ZZ\,$_{\mathrm{cand}}$ &  ZZCeti & --\\ 
SDSS\,J0158$-$0000 & 13,020 ? 440 & 8.29 ? 0.09 & ZZ\,$_{\mathrm{cand}}$ & -- & --\\ 
SDSS\,J0209+0050 & 24,910 ? 510 & 7.94 ? 0.07 & DA & -- & --\\ 
SDSS\,J0247+0003 & 19,362 ? 120 & 8.05 ? 0.019 & DA & -- & --\\ 
SDSS\,J0318+0030 & 11,450 ? 120 & 8.33 ? 0.05 & ZZ\,$_{\mathrm{cand}}$ & ZZ Ceti & ZZ Ceti$^b$\\ 
SDSS\,J0321$-$0050 & 16,896 ? 440^e & 8.00 ? 0.09^e & DAH & -- & --\\
SDSS\,J0326+0002 & 21,710 ? 320 & 7.93 ? 0.05 & DA & -- & --\\ 
SDSS\,J0326+0018 & 12,570 ? 90 & 8.17 ? 0.03 & ZZ\,$_{\mathrm{cand}}$ & -- & NOV0.5\,\%$^b$\\ 
SDSS\,J0342+0024$\dagger$ & 32,812 ? 170^e & 8.52 ? 0.03^e & DAB & -- & contam.\,(Sect.\ref{cont_st82})\\ 
SDSS\,J0349$-$0059  & 100,000 ? 790^e & 5.00 ? 0.01^e & PG1159 & -- & known pulsator$^f$\\
SDSS\,J2109+0111 & 20,530 ? 340^e & 7.81 ? 0.05^e & DA & -- & --\\
SDSS\,J2156$-$0046  & 10,940 ? 150 & 8.20 ? 0.09 & ZZ\,$_{\mathrm{cand}}$ & ZZ Ceti & --\\
SDSS\,J2157+0037 &  &  & no spec & -- & see Sect.\,\ref{phot_cand_st82}\\
SDSS\,J2157$-$0044$\dagger$ & 60,420 ? 5190 & 8.00 ? 0.31 & DA & -- & pos. err (Sect. \ref{err_pos})\\ 
SDSS\,J2218$-$0000 & 12,239 ? 440^e & 9.73 ? 0.24^e & MWD & -- & --\\
SDSS\,J2220$-$0041 & 7730 	? 40 & 7.98 ? 0.07 & DA & -- & WD+BD$^g$\\ 
SDSS\,J2237$-$0101 & 11,700 ? 270 & 8.06 ? 0.11 & ZZ\,$_{\mathrm{cand}}$ & ZZ Ceti & --\\ 
SDSS\,J2318$-$0114 & 21,470 ? 790 & 7.83 ? 0.11 & DA & -- & --\\ 
SDSS\,J2330+0100 & 6660 	? 60 & 8.23 ? 0.13 & DA & -- & --\\ 
SDSS\,J2333+0051 & 22,857 ? 890^e & 8.09 ? 0.07^e & V777\,Her$_{\mathrm{cand}}$ & NOV2.1\,\% & --\\\\ 
\hdashline\\
SDSS\,J0327+0012 & 16,800 ? 220 & 8.59 ? 0.03 & DA & NOV1.2\,\% \\ 		 	
SDSS\,J2245$-$0040 & 13,260 ? 300 & 8.12 ? 0.07 & DA & NOV4.2\,\% & \\
SDSS\,J2336$-$0051 & 12,860 ? 330 & 7.80 ? 0.09 & DA & NOV3.0\,\% & NOV0.5\,\%$^b$\\ 
\hline
\hline\\ [-1.5ex]
\multicolumn{6}{l}{$^a$\,\citet{tremblayetal11-1};$^b$\,\citet{mukadametal04-1}; $^c$\,\citet{kleinmanetal04-1}; $^d$\,\citet{kepleretal15-1};}\\
\multicolumn{6}{l}{$^e$\,\citet{Kleinmanetal13-1}; $^f$\,\citet{woudtetal10-1}; $^g$\,\citet{steeleetal09-1}}

\end{tabular}
\end{table*}

\begin{figure*}
\centering
\includegraphics[width=\columnwidth ]{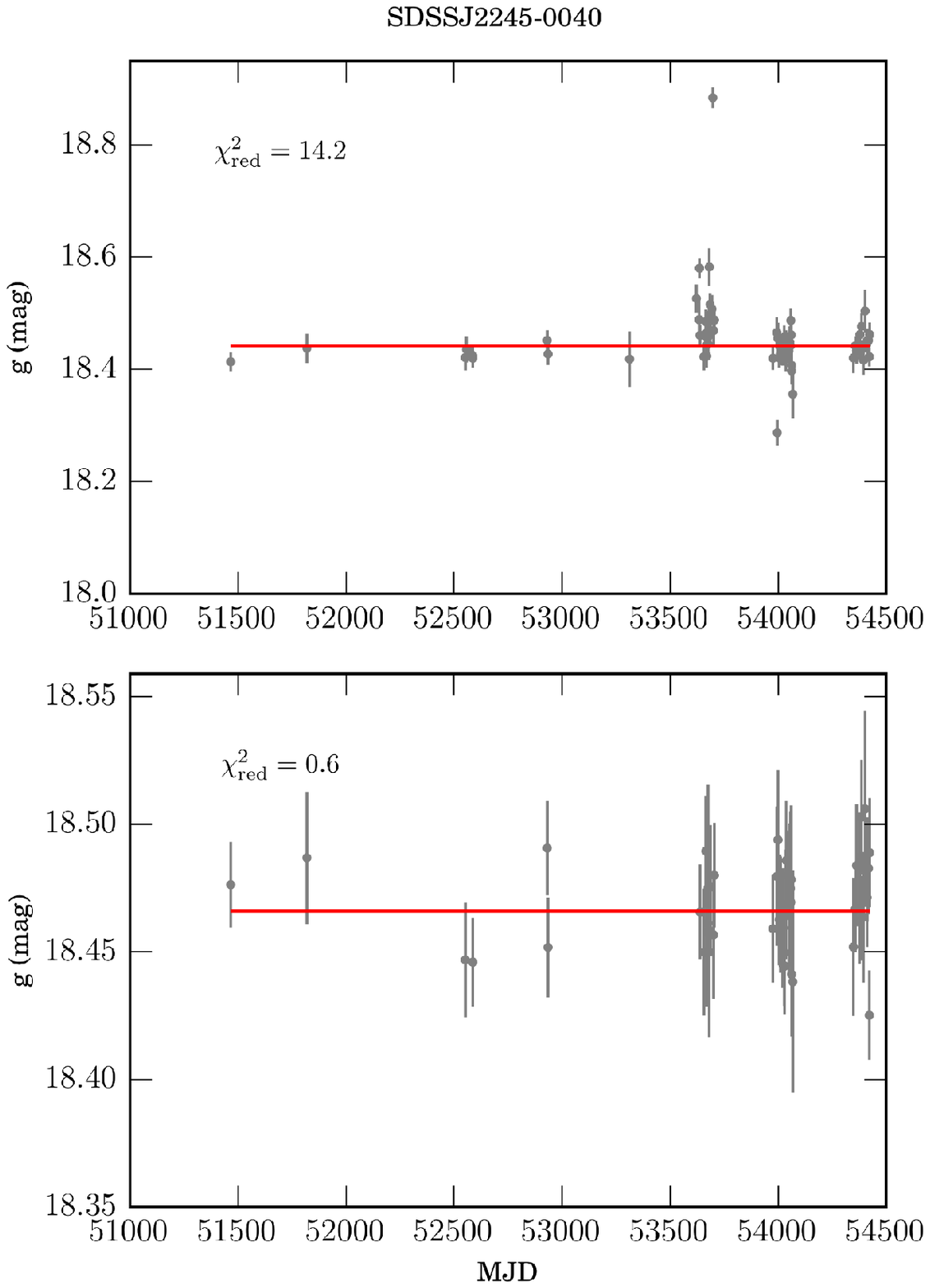}
\includegraphics[width=\columnwidth ]{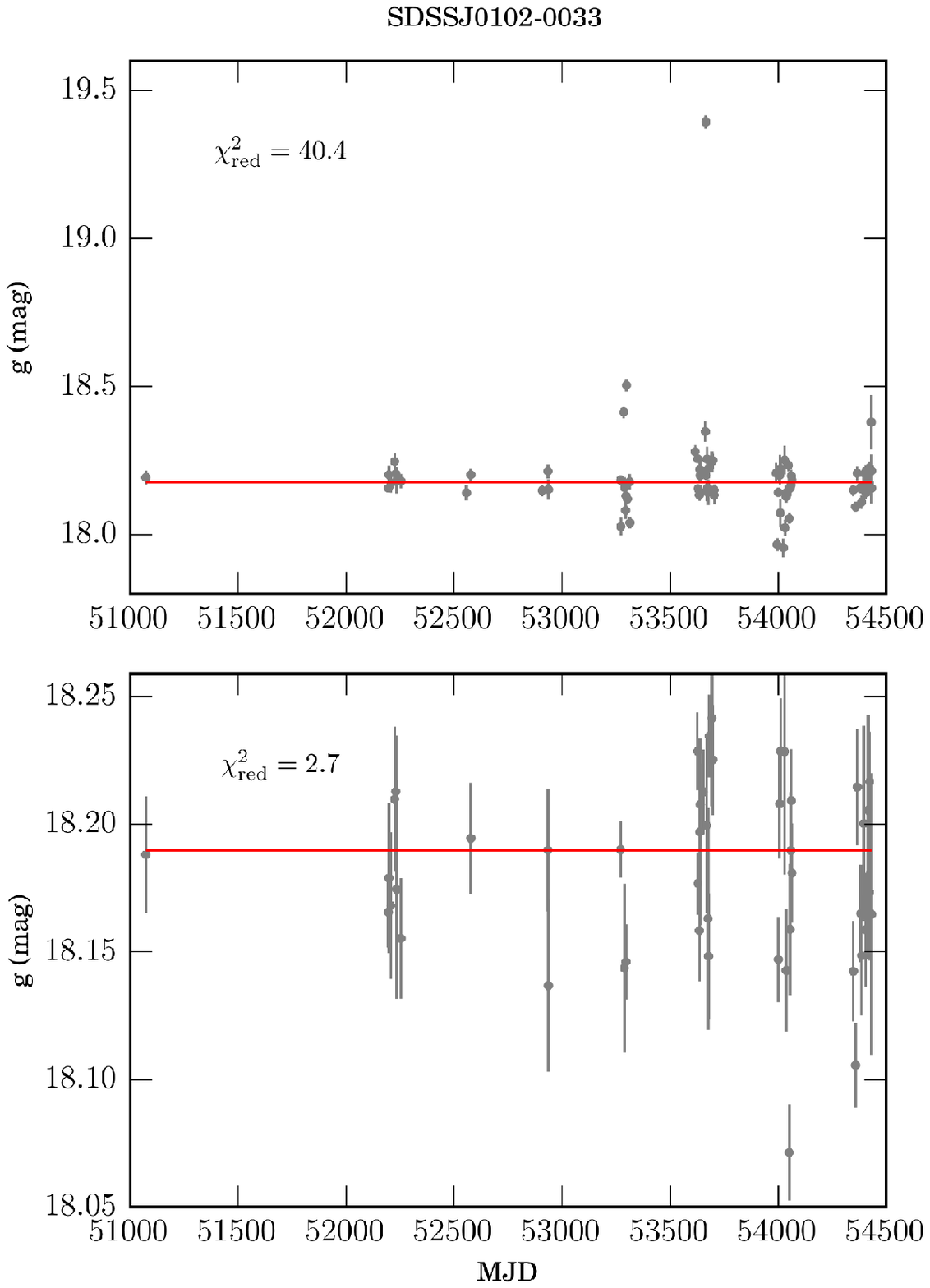}

\caption{\label{light_vs} Multi-epoch light curves of the ``control" white dwarf SDSS\,J2245$-$0044 (\emph{left}) and the confirmed ZZ Ceti SDSS\,J0102$-$0033 (\emph{right}).
\emph{Top panels:} all available Stripe\,82 data with the default calibration. \emph{Bottom panels:} Multi-epoch light curves of the same stars  after re-calibrating the photometry and discarding unreliable nights. The red line indicates the median magnitude values.}
\end{figure*}

Following  this strict selection criteria, 85 of our white dwarfs candidates were left with less than eight epochs of reliable data, which we deemed insufficient for a robust variability analysis. Consequently,  we dropped these objects, reducing our sample size to 315.
For each of these white dwarf candidates, we quantify the scatter of the recalibrated $g$-band  magnitudes with respect to  their median value (and therefore the degree of variability) by calculating reduced $\chi^2$ values (Fig. \ref{light_vs}).
We defined ${\chi^2}_{\mathrm{red}} >2.0$ as the threshold for selecting 26 variable white dwarf candidates.
For each of these stars, we constructed multi-epoch light curves, calculated amplitudes, and  retrieved SDSS images and spectra (where available; Table\,\ref{phot_par}).
The amplitude of the scatter in the data  was calculated from the observed light curves using a Monte Carlo method. We randomly varied the magnitude of each data point using a Gaussian probability distribution whose width was set to the uncertainty in the Stripe\,82 magnitudes. For each re-sampling we calculate the amplitude as half of the difference between the brightest and faintest detection. The reported amplitudes are the averages of 1000 re-samplings. These values reflect  the minimum expected amplitude of variability required to cause the observed scatter in the Stripe\,82 data.

All but one of our candidates (SDSSJ2157+0037, see Sect.\,\ref{phot_cand_st82}) have an SDSS spectrum.
Combining all available data (spectra, SDSS colours and images, ${\chi^2}_{\mathrm{red}}$ and light curves) we then attempted	 to assess the nature of the observed variability.

\subsection{Source contamination}\label{cont_st82}
\begin{figure}
\centering
\includegraphics[width=0.5\columnwidth ]{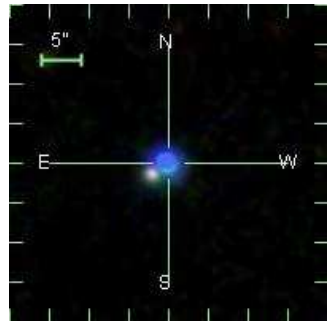}
\caption{\label{SDSSJ0342+0024.eps} SDSS image of SDSS\,J0342+0024 showing possible contamination from nearby source.}
\end{figure}

If two sources in Stripe\,82 happen to be spatially very close ($\lesssim 2"$), some of the measurements may suffer from poor deblending during variable seeing conditions, resulting in apparently variable multi-epoch photometry.
Inspecting the SDSS images of the 26 variable white dwarf candidates, we found that  SDSS\,J0342+0024 has a close, potentially contaminating neighbour (Fig. \ref{SDSSJ0342+0024.eps}). We therefore decided to drop this star from our candidate list.

\subsection{Errors in source extraction}
\label{err_pos}
While inspecting the light curves of our variable candidates, two objects stood out for having some extremely faint detections.
SDSS\,J0106$-$0014 has two $g$-band magnitude measurements of 25.5 mag and 24.6 mag, while SDSS\,J2157$-$0044 has one $g$-band magnitude measurement of  24.7 mag. All these values are much fainter than the nominal $g$-band limit of SDSS, implying that the objects were most likely just at the edge of detection in those nights.
In both cases these magnitudes diverge dramatically from the median magnitude of the object ($g=18.63$ for SDSS\,J2157$-$0044, $g=18.15$ for SDSS\,J0106$-$0014), strongly contributing to the high reduced $\chi^2$ values calculated for these objects. We decided to further verify the reliability of these measurements by checking the coordinates of the detections.

\begin{figure}
\hspace{0.15cm}
\includegraphics[width=\columnwidth]{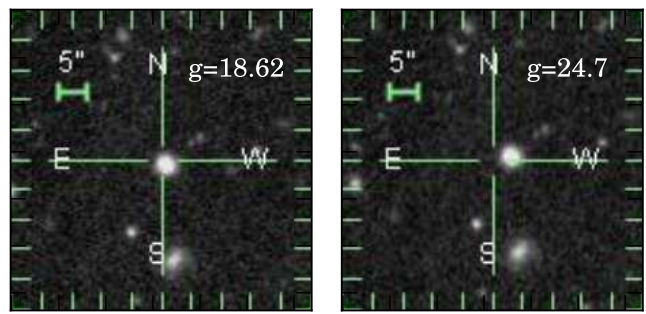}
\caption{\label{SDSS2157-0044} SDSS images of SDSSJ\,2157$-$0044 centred at the position of one of the detections close to the median magnitude (\emph{left panel}) and at the position of the faint detection (\emph{right panel}).}
\end{figure}

Figure\,\ref{SDSS2157-0044} clearly shows that, in the case SDSS\,J2157$-$0044, the apparent dimming is due to an error in the position at which the source was extracted. 
However for SDSS\,J0106$-$0014 the coordinates of bright and faint detections are consistent, and we have to conclude that a genuine dimming was observed (see Sect.\,\ref{eclipse}). 

\begin{figure}
\includegraphics[width=\columnwidth]{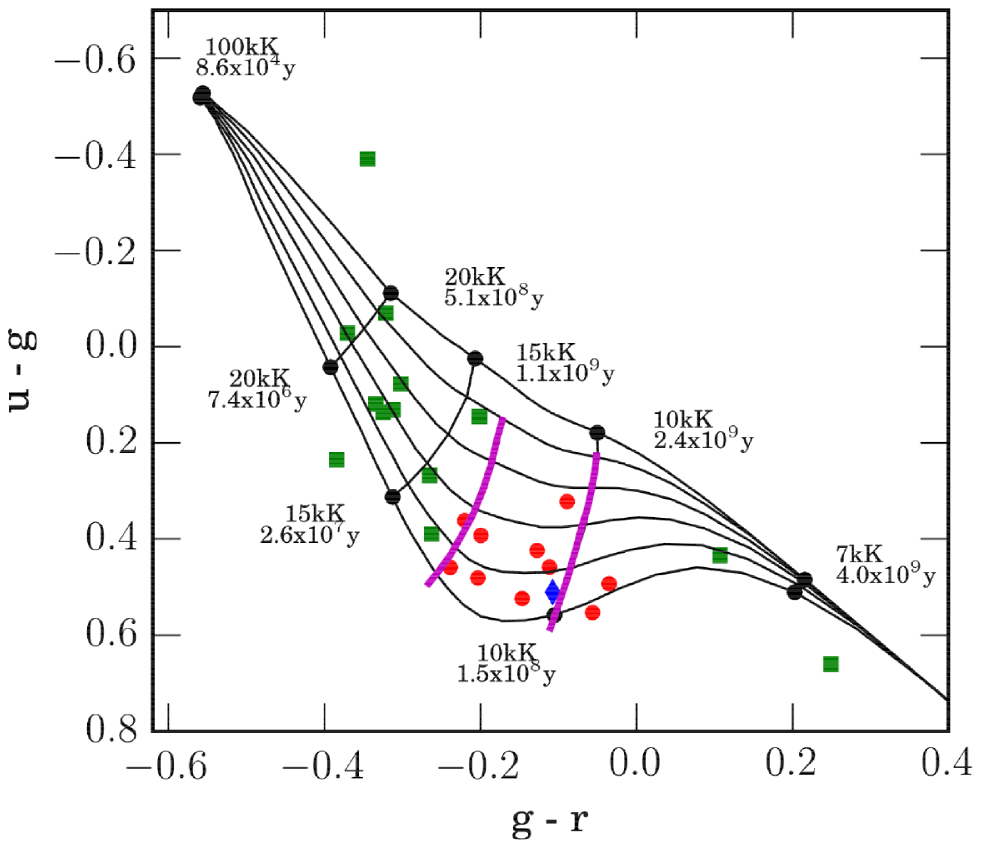}
\caption{\label{colour-comp} $u-g,g-r$ colour-colour distribution of our sample of Stripe\,82 variable white dwarfs candidates: ZZ Ceti candidates as red circles, one candidate with no SDSS spectroscopy (see Sect.\,\ref{phot_cand_st82}) as the blue diamond and remaining candidates as green squares. The magenta lines show the empirical boundary of the ZZ Ceti instability strip from \citet{gianninasetal14-1}. White dwarf cooling tracks from \citet{holberg+bergeron06-1} are shown in the overlay. The colours of our variable candidates have been computed using the median magnitude values.}
\end{figure}

\section{ZZ Cetis candidates}\label{zz_c}
Since the ZZ Ceti instability strip is defined by a narrow  range in temperature and surface gravity, a $u-g, g-r$ colour projection of the $T_{\mathrm{eff}}/ \log g$ strip can be used to select ZZ Ceti candidates \citep{mukadametal04-1, greissetal14-1}.
By inspecting the colour-colour locus of known ZZ\,Cetis we selected from our variable candidate list ten stars most likely belonging to this class (Fig.\,\ref{colour-comp}, Table\,\ref{lit_par}). 
However, as we discuss later, $u-g, g-r$ colours alone are not very efficient at discriminating between ZZ Ceti and non pulsating white dwarfs (see Sect.\,\ref{istrip}).

\subsection{Follow up observations}\label{lt_follow}
We obtained high-speed photometry for six of our ten ZZ Ceti candidates and one V777\,Her  candidate (see Sect.\,\ref{other_pulse_st82}) in order to confirm their pulsating nature. 
We used the optical imaging component of the IO (Infrared-Optical) suite of instruments (IO:O) \footnote{http://telescope.livjm.ac.uk/TelInst/Inst/IOO/} on the Liverpool Telescope (LT) on the island of La Palma. Each target was observed with 30s exposures for  $\simeq 2$ hours.
In order to verify the robustness of our multi-epoch variability selection we also observed three ``control" white dwarfs which have colours compatible with those of ZZ Cetis, but for which we found no evidence of variability in multi-epoch data (i.e. $\chi^{2}_{\mathrm{red}} \leq1.0$; Table\,\ref{phot_par}).

\begin{figure*}
\includegraphics[width=2\columnwidth ]{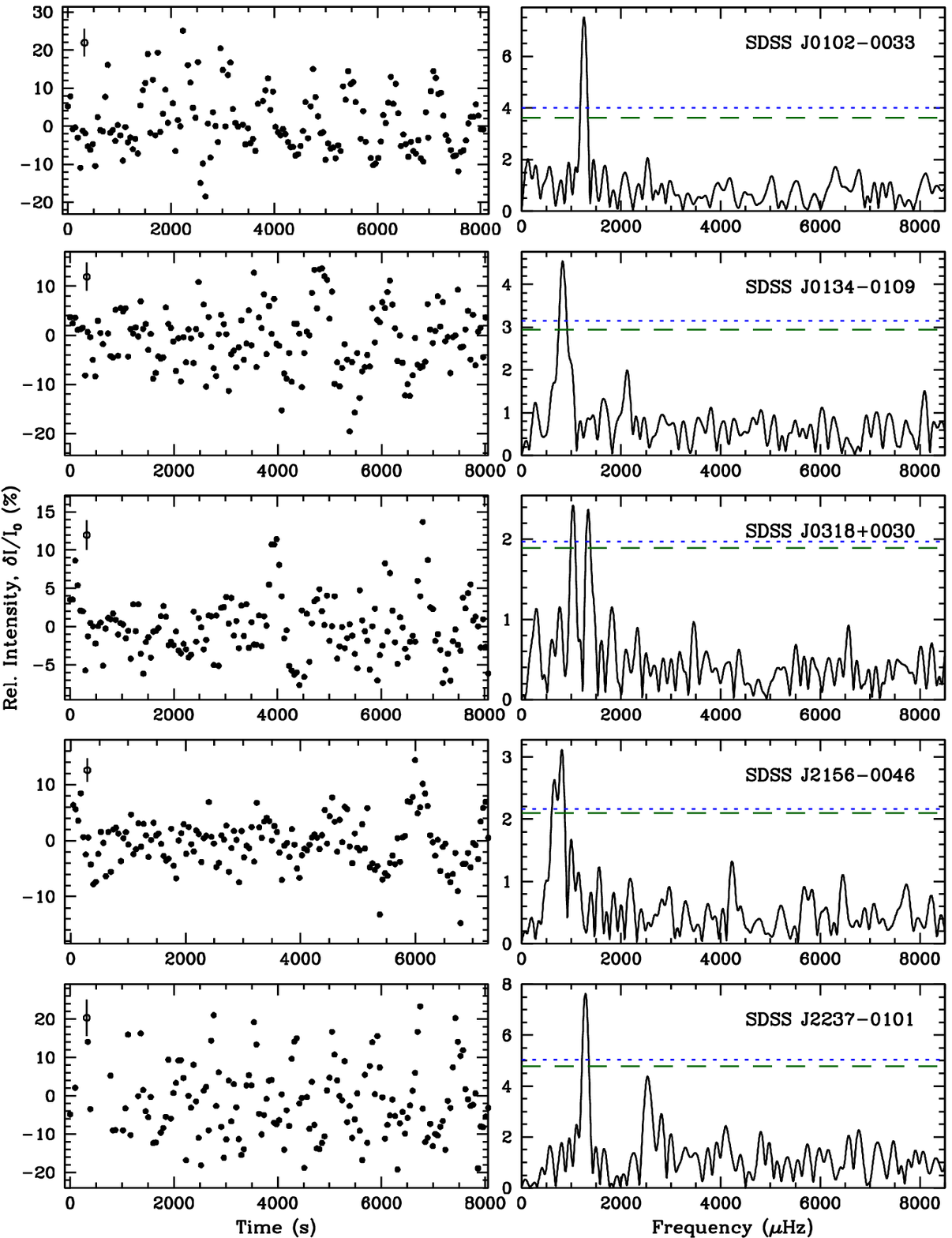}
\caption{\label{pulse} LT ligthcurves (left) and corresponding Fourier transforms (right) for the five ZZ Ceti candidates which where confirmed as pulsating white dwarfs. Typical photometric uncertainties are indicated in the top left of each panel. The green dashed line marks 4 times
the average amplitude of the entire Fourier transform; the blue dotted line marks the 3-sigma significance threshold, as determined by 10,000 bootstrap shuffles of the light curve (for more details see \citealt{greissetal14-1}). SDSS\,J0134$-$0109, SDSS\,J2237$-$0101 and  SDSS\,J2156$-$0046 are new high-amplitude ZZ Ceti discoveries.}
\end{figure*}

\begin{figure*}
\includegraphics[width=2\columnwidth ]{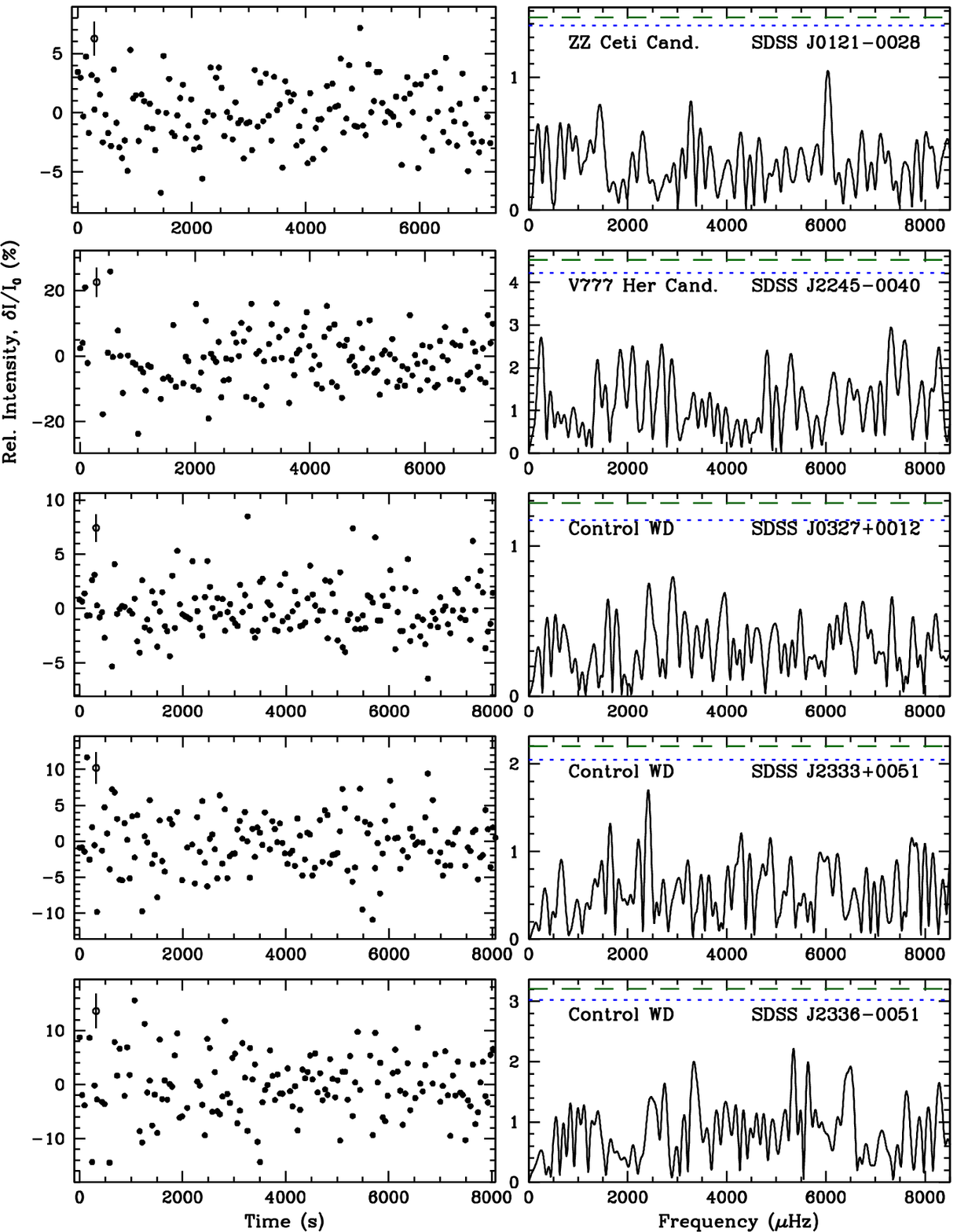}
\caption{\label{non_pulse} As Figure\,\ref{pulse} for the five objects for which we detected no variability in our LT observations. SDSS\,J0327+0012, SDSS\,J2245$-$0040 and SDSS\,J2336$-$0051 have colours compatible with those of ZZ Cetis, but show no evidence of variability in their Stripe\,82 data (``control" white dwarfs). Our observations do not completely rule out the presence of pulsations, as all objects may vary at smaller amplitudes; the maximum pulsation amplitude of known ZZ\,Cetis has a median of 1.5 per cent in the SDSS $g$ band.}
\end{figure*}

\subsection{Analysis and results: three new ZZ Cetis}
\label{zz_confirm}
We extracted sky-subtracted light curves from our LT observations and computed Fourier transforms (FT) for our six ZZ Ceti candidates, the three ``control" DAs and one candidate V777\,Her (see Sect.\,\ref{other_pulse_st82}).
We classified a candidate as a confirmed pulsating white dwarf if its FT shows a peak larger than 4 times the average amplitude of the entire FT and above a 3\,$\sigma$ threshold \citep{greissetal14-1}. 
Of the six ZZ Ceti candidates observed we detect pulsations for five of them. Two (SDSS\,J0102$-$0033 and SDSS\,J0318+0030) had already been identified as ZZ Ceti stars by \citet{mukadametal04-1}, while the remaining three (SDSS\,J0124$-$0109, SDSS\,J2157$-$0044 and SDSS\,J2237$-$0101 \footnote{In the referee report, it was brought to our attention that SDSS\,J2237$-$0101 had been independently identified as a ZZ\,Ceti by Wolf et al. (2007, private communication) who selected SDSS\,J2237$-$0101 from a small sample of white dwarfs with SDSS spectroscopy which appeared in the preliminary Stripe\,82 variability catalogue of \citet{ivezicetal07-1}.}) are new discoveries (Fig.\,\ref{pulse}).
We designate the remaining five objects (three ``control" white dwarfs, one ZZ Ceti candidate and one V777\,Her candidate) as ``not observed to vary" (NOV) and estimate percentile non variability limits (Fig.\,\ref{non_pulse}).
These results are reported in Table \ref{lit_par}.

\subsection{The ZZ Ceti instability strip}
\label{istrip}
\begin{figure}
\includegraphics[trim={19pt 0 5pt 0}, width=\columnwidth]{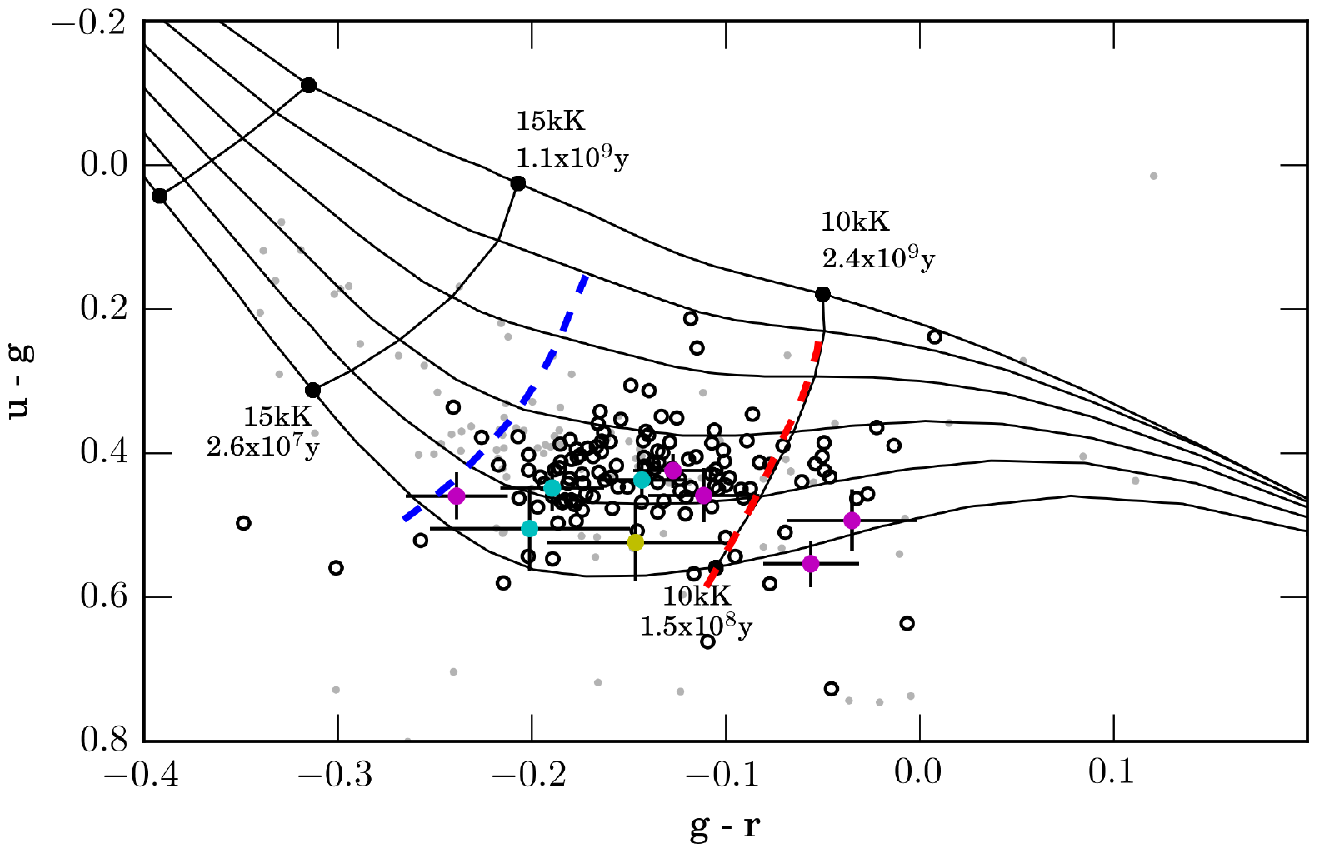}
\includegraphics[width=\columnwidth]{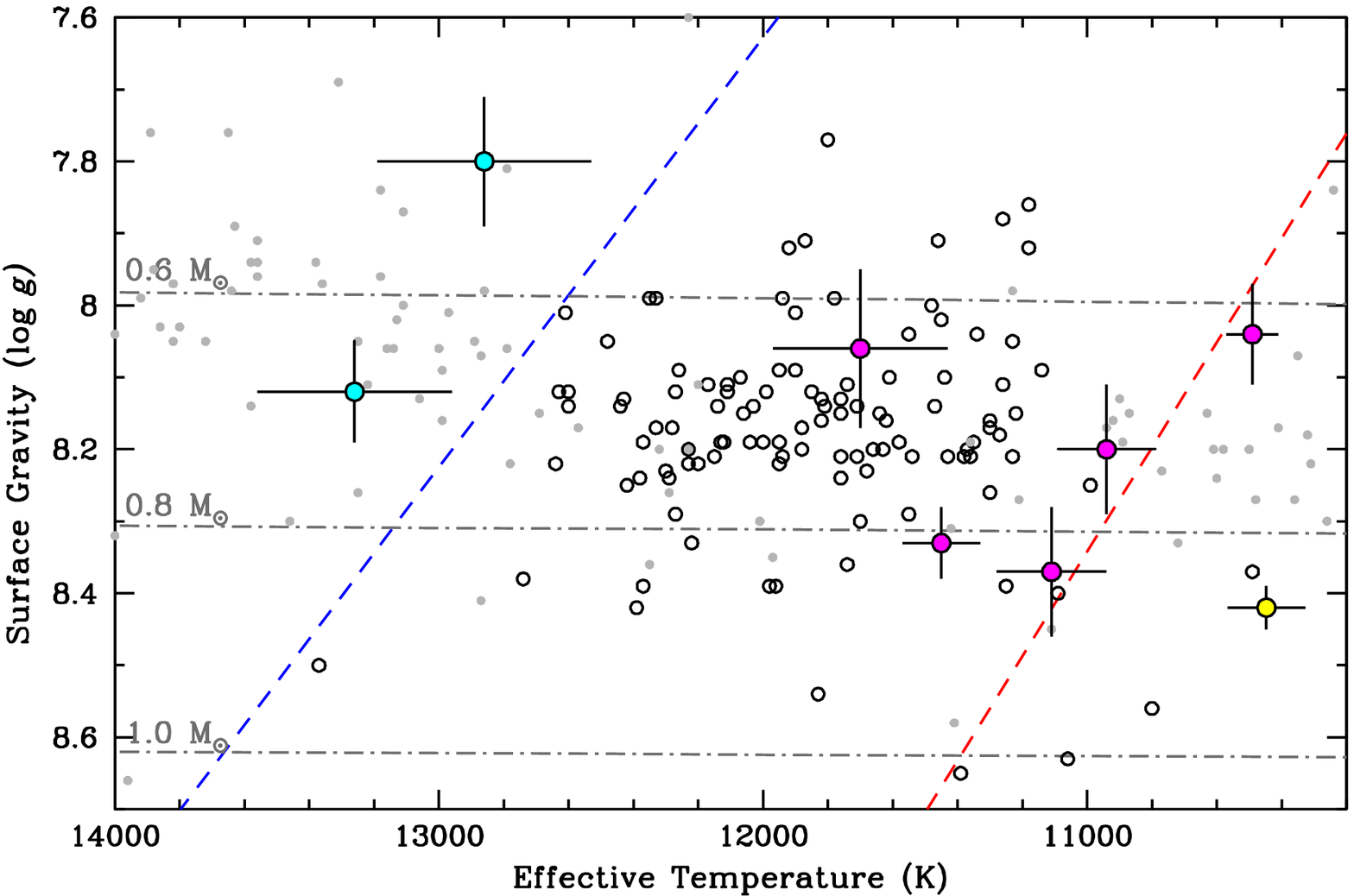}

\caption{\label{good_istrip}
Colour distribution (\textit{top panel}), and $T_{\mathrm{eff}}$-$\log g$ distribution (\textit{bottom panel}) of known ZZ\,Cetis (empty circles) and NOVs (grey dots). $\log g$ and $T_{\mathrm{eff}}$ values were  taken from \citet{tremblayetal11-1} and \citet{gianninasetal11-1} and are calculated using 1D atmospheric models. The empirical blue and red edge of the ZZ Ceti instability strip are shown as a blue and a red dashed line, respectively. We include our confirmed ZZ Cetis as magenta dots and our ``control"  white dwarf confirmed as NOVs as cyan dots (Table\,\ref{lit_par}) and the ZZ Ceti candidate SDSS\,J0121$-$0028 (Sect.\,\ref{IR_obj}) as a yellow dot. The ``control"  white dwarf SDSS\,J0327+0012 is outside the range of the plot in the bottom diagram. }
\end{figure}
\begin{figure*}
\includegraphics[width=2\columnwidth]{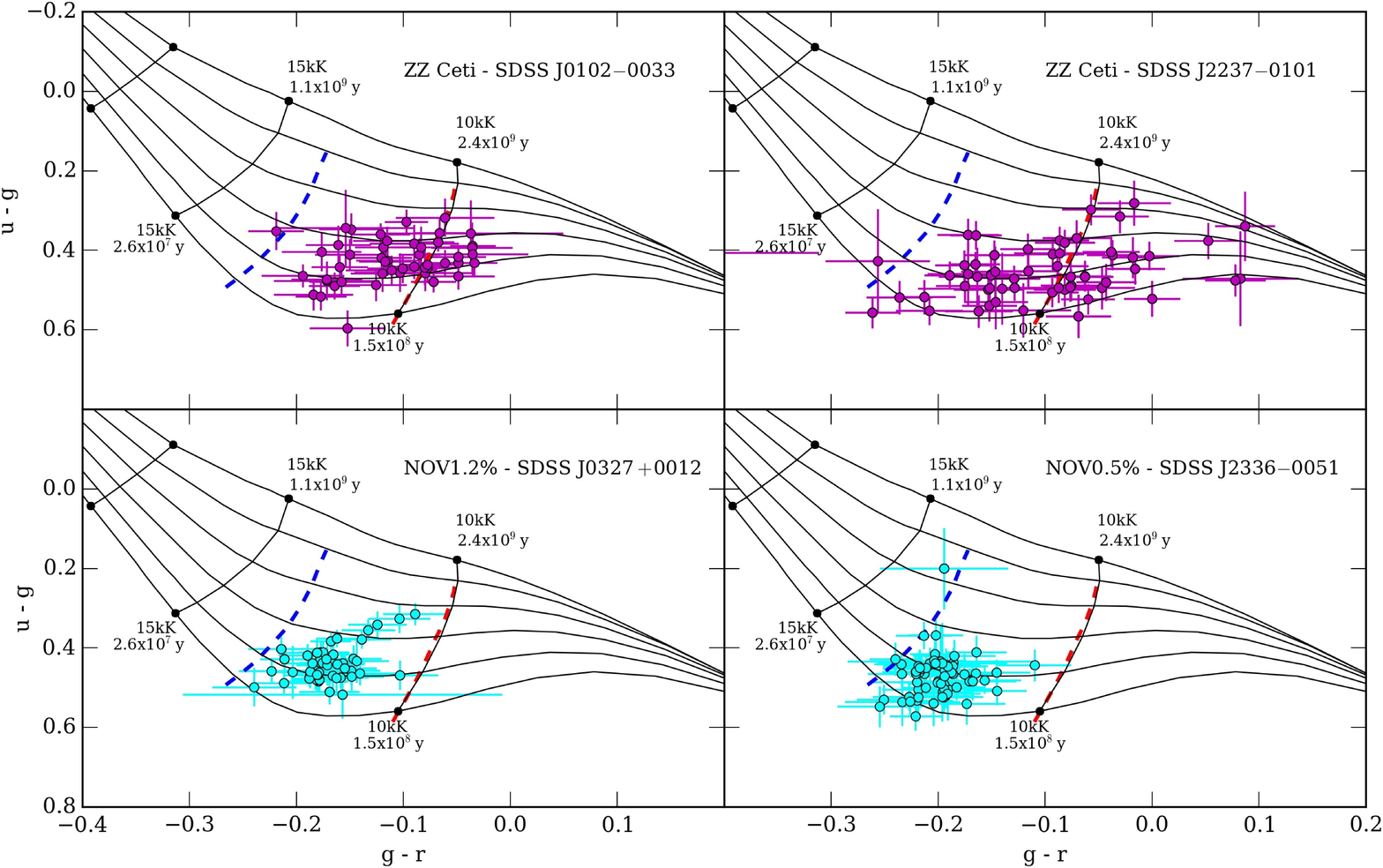}
\caption{\label{colour_scatter} Colour-colour distribution of all reliable recalibrated Stripe\,82 epochs of two confirmed ZZ Cetis (SDSS\,J0102$-$0033, SDSS\,J2237$-$0101, \emph{top panels}, magenta points) and two confirmed NOV, ``control" white dwarfs  (SDSS\,J2245$-$0040 and SDSS\,J2336$-$0051, \emph{bottom panels}, cyan points).
The empirical blue and red edge of the ZZ Ceti instability strip are shown as a blue and a red dashed line, respectively.}
\end{figure*}

As mentioned in Section \ref{zz_c} we picked  our ZZ\,Ceti 
candidates by selecting the Stripe\,82 variables which were closest to the $u-g,g-r$ projection of the ZZ\,Ceti instability strip.
However, Fig.\,\ref{good_istrip} reveals that many known NOVs are within the $u-g,g-r$ boundaries used; and vice versa, confirmed ZZ\,Cetis lie significantly outside the strip. 
Several studies have shown that a candidate selection purely based on colours yields a success rate of only $13 - 30$ per cent (see \citealt{vossetal06-1, fontaineetal82-1, mukadametal04-1}). Even though $u-g,g-r$ colours are good indicators of white dwarf temperature it appears that the narrow range of $T_{\mathrm{eff}}$ and  $\log g$ that defines the ZZ Ceti instability strip does not unambiguously project to the observed colours.
Furthermore,  the $ugriz$ magnitudes of SDSS are not acquired simultaneously, but in the sequence $riuzg$, with observations in each filter separated by 71.72 s.
As a result $u$ and $g$ band observations are performed roughly 143 seconds apart and $g$ and $r$ observation nearly 290 seconds apart. Cool pulsating white dwarfs can vary in brightness by up to 20 percent on such timescales, implying that the SDSS colours of these objects may be taken during different pulsation phases and are not, therefore, reliable indicators of temperature. Figure\,\ref{colour_scatter} clearly illustrates this effect, i.e. multi-epoch colours of confirmed ZZ\,Cetis exhibit much larger scatter in colour-colour space than those of confirmed NOVs. Reddening too, though often considered to be a minor effect for relatively nearby white dwarfs, may still contribute to the observed discrepancy between temperatures and colours.

If optical spectra are available, $T_{\mathrm{eff}}$  and $\log g$ can be measured from fitting the Balmer lines of DA white dwarfs. (\citealt{koesteretal79-1}, \citealt{bergeronetal95-1}, \citealt{koester09-1}, \citealt{tremblayetal11-1}).  ZZ Ceti candidates can then be reliably selected on the basis of the atmospheric parameters (\citealt{bergeron01-1},  \citealt{mukadametal04-1},  \citealt{gianninasetal11-1}). 
Figure\,\ref{good_istrip} illustrate that the majority of known ZZ\,Cetis (including the five confirmed as part of this work) lie within the boundaries of the empirical instability strip, with little contamination from NOVs. 
Even though the spectroscopic method can achieve efficiency upwards of 80 per cent, the required large samples of spectroscopically confirmed DAs are observationally expensive.

In this pilot study we selected ZZ\,Ceti candidates without relying on spectroscopy and combining colour selection with the evidence of multi-epoch variability. Five out of the six candidates followed-up with high speed photometry,  were confirmed to be ZZ Cetis (see Sect.\,\ref{zz_confirm}), implying an efficiency  $\simeq83$ per cent. 
Even though limited to sparse data over only the  300\,deg$^2$ of Stripe\,82, we showed that, using multi-epoch data, it is possible to achieve an efficiency similar to that of the spectroscopic selection method. However it is important to keep in  mind that our statistics is limited to only the six ZZ Ceti candidates we followed up with LT observations. 
Our selection method is biased in favour of cool, high-amplitude ZZ Cetis (see Sect.\,\ref{pulse_prop_sect}) and our three	 newly identified ZZ Cetis are among the coolest ever discovered. With their identification we empirically  constrain  the red edge of the instability strip.
Applying this selection to other current and future time domain surveys (e.g. PanSTARSS, LSST) will provide very large samples of high-confidence ZZ Ceti candidates, paving the way for global ensemble asteroseismology of white dwarfs.


\subsection{Pulsation properties of the ZZ Ceti variables}
\label{pulse_prop_sect}
\begin{table}
\centering
\caption{\label{pulse_prop}  Pulsation properties of the five confirmed ZZ Cetis. Three new ZZ Ceti discovered as part of this work are marked with *.}
\begin{tabular}{l D{?}{\pm}{3.3} D{?}{\pm}{3.3}}
\hline
name & \multicolumn{1}{l}{period (s)}& \multicolumn{1}{l}{amplitude (mma)}\\
\hline
\hline\\ [-1.5ex]
SDSS\,J0102$-$0033 & 796.1?3.7 & 75.1?6.5\\
SDSS\,J0134$-$0109* & 1212?13 & 45.4?5.9\\
SDSS\,J0318+0030 & 969?11 & 21.8?3.5\\
& 746.4?6.7 & 21.1?3.5\\
SDSS\,J2156$-$0046* & 1234?15 & 31.4?4.1\\
& 1478?21 & 27.0?3.5\\
SDSS\,J2237$-$0101* & 774.4?4.6 & 80.1?8.5\\
& 392.3?2.2 & 44.7?8.3\\

\hline
\hline
\end{tabular}
\end{table}

We list the dominant periods found for the five confirmed ZZ Cetis in Table\,\ref{pulse_prop}. All five ZZ Cetis undergo large amplitude pulsation with periods longer than 600s, which are normally associated with cool ZZ Cetis ($T_{\mathrm{eff}} \simeq 11,000$\,K). Fig.\,\ref{good_istrip} clearly shows that all our confirmed ZZ Cetis are indeed cool pulsators, with three of them  lying on the red edge of the instability strip. This selection effect is caused by the fact that pulsation amplitude increases with decreasing $T_{\mathrm{eff}}$. Since variability with amplitudes $\lesssim$0.03 mag would  not be detectable in the Stripe\,82 data, we are biased to preferentially select cool, large-amplitude ZZ Cetis.

\citet{mukadametal04-1} report periods for SDSS\,J0102$-$0033 (926.1\,s, 830.3\,s) and  SDSS\,J0318+0030 (826.4\,s, 587.1\,s, 536.1\,s) compatible with cool ZZ Cetis, but significantly different from the period reported here. Such changes in pulsation periods over long time scales are not uncommon, as cool ZZ Cetis are known to undergo amplitude and frequency variations (e.g. G29$-$38, \citealt{mcgrawetal75-1, kleinmanetal98-1}; GD 1212, \citealt{hermesetal14-2}). 

\section{Notes on single objects}
\subsection{SDSS\,J0121$-$0028}
\label{IR_obj}
Based on its $u-g; g-r$ colours, SDSS\,J0121$-$0028 was selected as a ZZ Ceti candidate, but in our two hours high-speed photometric follow-up we did not observe any pulsation to a limit of 1.39 per\,cent. 
The amplitude of the variation calculated from the Stripe\,82 data (8-10 per cent, Table\,\ref{phot_par}) is above this non-variability threshold, possibly implying that our LT observations were taken during a
period of destructive interference of the pulsation modes
(e.g. \citealt{castanheiraetal06-1}). Nonetheless, Fig.\,\ref{good_istrip} shows that SDSS\,J0121$-$0028 lies outside the ZZ Ceti instability strip, red-ward of the the red edge. 
If SDSS\,J0121$-$0028 is indeed a ZZ Ceti it would be a rare, though not unprecedented, outlier (e.g. WD\,J0940+0050, \citealt{castanheiraetal13-1}). 
Another possibility is that  SDSS\,J0121$-$0028 undergoes some other type of magnitude variation on timescales longer than 2 hours. 
\begin{figure*}
\subfigure{\includegraphics[trim={20pt 0 0 5pt},width=\columnwidth]{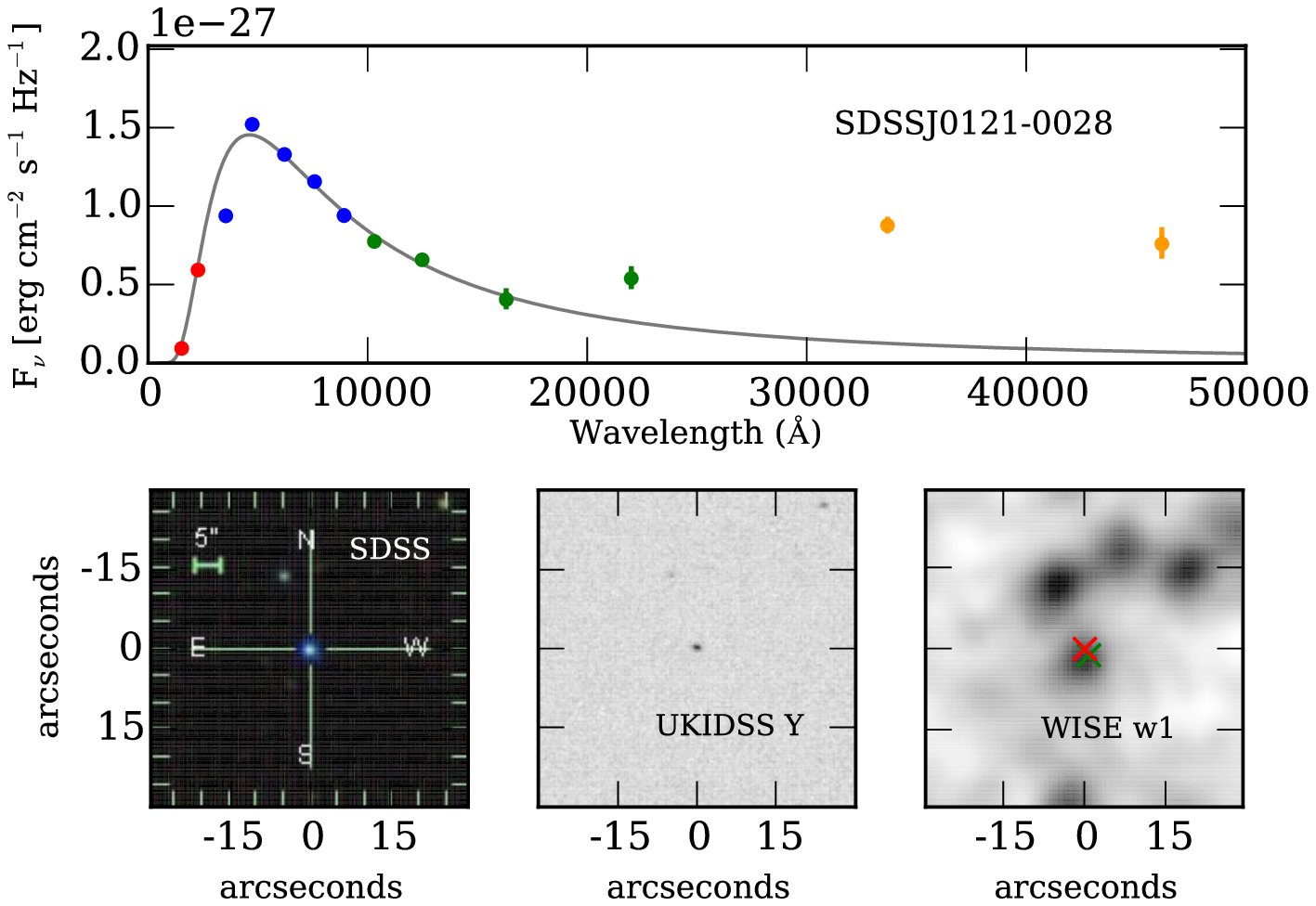}}
\subfigure{\includegraphics[width=\columnwidth]{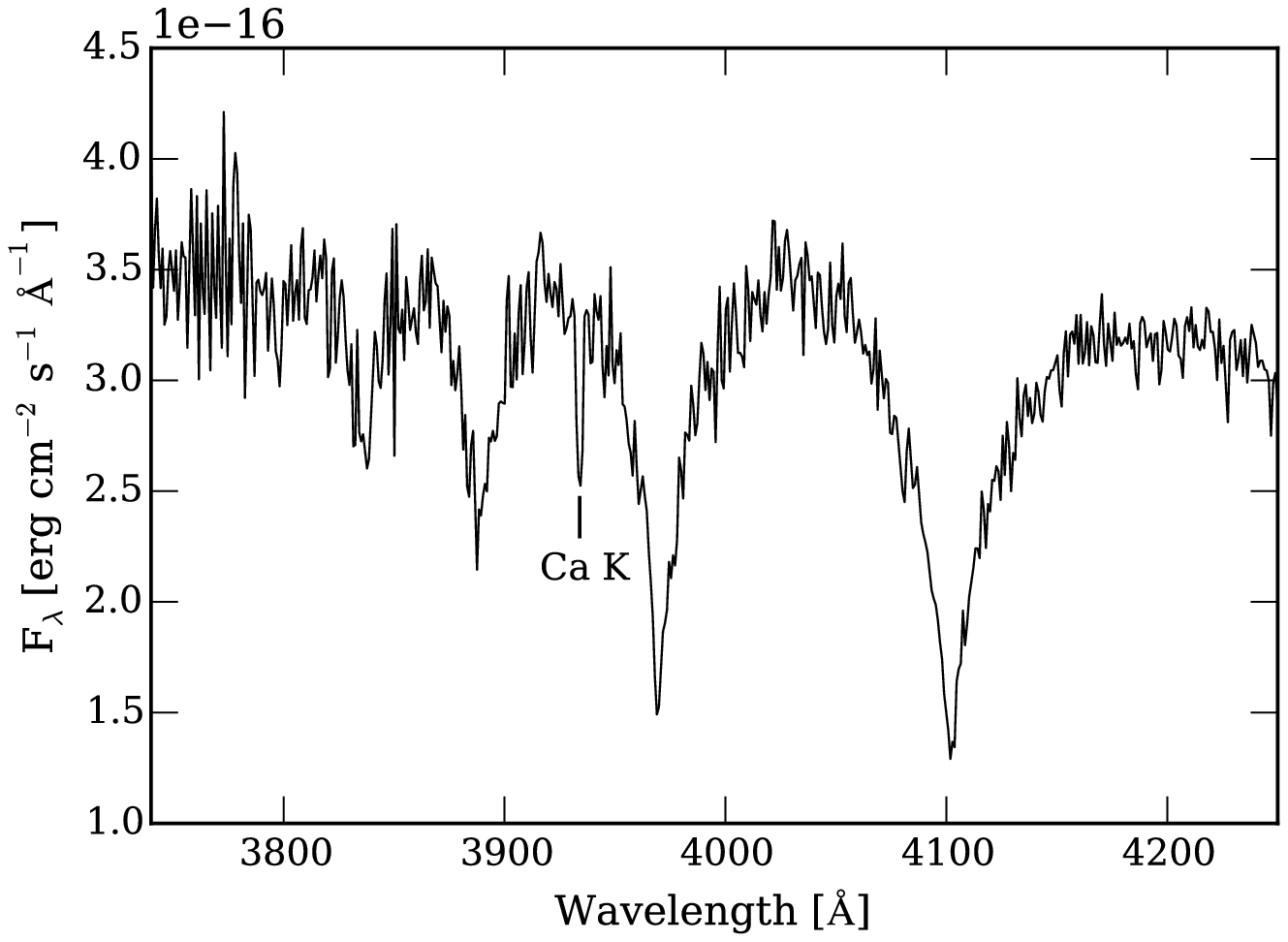}}

\caption{\label{SED}  \textit{Left panel}: Spectral energy distribution of SDSS\,J0121$-$0028. Galex $fuv, nuv$ fluxes are plotted in red, SDSS Stripe\,82 recalibrated median  $ugriz$ fluxes in blue, UKIDSS $YJHK$ fluxes in green and WISE $w1, w2$ fluxes in yellow. The grey solid line shows a blackbody model of the white dwarf fitted to the $g$ and $r$ fluxes.
\textit{Right panel}: SDSS spectrum of SDSS\,J0121$-$0028 showing the presence of Ca K absorption line. }
\end{figure*}

To further investigate the nature of SDSS\,J0121$-$0028, we retrieved all available ultraviolet and  mid-infrared photometry of SDSS\,J0121$-$0028. The spectral energy distribution of SDSS\,J0121$-$0028 (Fig.\,\ref{SED}) shows a marked infrared excess which is not consistent with a single, isolated white dwarf.
The presence of a close, low-mass companion could explain the observed infrared excess and companions may also cause some degree of optical variability in white dwarfs (\citealt{littlefairetal14-1, casewelletal15-1, maozetal15-1}).

However the infrared excess in  SDSS\,J0121$-$0028 becomes apparent only in the  UKIDSS $K$ band, rising towards longer wavelengths. A blackbody fit to this infrared emission, adopting the photometric distance of the white dwarf, suggests a $T_{\mathrm{eff}} \simeq 1100$\,K and radius $\simeq 2.1$ Jupiter radii (R$_{\mathrm{J}}$), i.e. more than twice the radius of a typical brown dwarf (0.83\,R$_{\mathrm{J}}$, \citealt{sorahanaetal13-1}). It therefore seems unlikely that the infrared excess is caused by a brown dwarf companion.
An alternative origin of the infrared excess could be a circumstellar debris discs resulting from tidal disruption of rocky planetesimals (\citealt{grahametal90-1, jura03-1}), and  $\simeq 1-3.5$ per cent of all white dwarfs exhibit infrared excesses consistent with the presence of such discs \citep{farihietal09-1, girvenetal11-1, rocchettoetal15-1}. In all cases the presence of these discs is accompanied with metal pollution of the white dwarf atmosphere. 
Close inspection of the SDSS spectrum of SDSS\,J0121$-$0028 reveals a strong calcium absorption line at $3933.7 \mathrm{\AA}$, identifying it as a metal polluted DAZ white dwarf (Fig.\,\ref{SED}). The presence of metals in the photosphere of SDSS\,J0121$-$0028 strongly supports the hypothesis that the observed infrared excess is due to a debris disc.
There is growing evidence of variability in some white dwarf debris discs, including changes in the optical line profiles 
\citep{gaensickeetal08-1, wilsonetal15-1} and line fluxes \citep{wilsonetal14-1} from gaseous discs, as well as changes in the infrared flux from the dust \citep{xuetal14-1}. However, to date, optical variability in white dwarfs with debris discs has only been observed at amplitudes much lower than what we measured for SDSS\,J0121$-$0028.\\
The data at hand does not allow to unambiguously determine the nature of the variability of SDSS\,J0121$-$0028 and we recommend further observations of this object.

\subsection{SDSS\,J0050$-$0023 and SDSS\,J0326+0018}
Two of our ZZ Ceti candidates  had already been observed by \citet{mukadametal04-1} and found not to vary: SDSS\,J0050$-$0023 and SDSS\,J0326+0018. We did not acquire more observations, but the NOV limits calculated by \citet{mukadametal04-1} are considerably smaller than the amplitudes of the variability detected in the Stripe\,82 data  ($>$ 0.1 mag; Table\,\ref{phot_par}).
Again this implies that these white dwarfs may vary on longer timescales.

SDSS\,J0050$-$0023 is known massive white dwarf with spectroscopic mass above a 1$M_{\mathrm{\odot}}$ \citep{castanheiraetal10-1}. At the temperatures around the ZZ Ceti instability strip, stars with such mass are expected to be up to 90 per cent
crystallized \citep{kanaanetal05-1}. Crystallization has significant effect on the pulsation properties of a white dwarfs since pulsations cannot propagate into the crystallized region \citep{montgomery+winget99-1}. Pulsating massive white dwarfs can therefore experience dramatic changes in their pulsation amplitudes (e.g. BPM37093, \citealt{mcgraw76-1, kanaanetal98-1}).

SDSS\,J0050$-$0023 was classified as  NOV6 by \citet{mukadametal04-1} and NOV3.7 by \citet{castanheiraetal10-1}, however \citet{castanheiraetal10-1} also report possible, lower-amplitude pulsations with a period of 584s. The multi-epoch variability observed in Stripe\,82 seems to validate the hypothesis that SDSS\,J0050$-$0023 is a massive pulsator with highly variable pulsation amplitudes. It is possible that all high-cadence monitoring to date was carried out during a phases of pulsation dampening and consequently failed to identify pulsations. 
We encourage further, long-term monitoring of SDSS\,J0050$-$0023 to determine if the star is truly variable. 


\subsection{SDSS\,J0106$-$0014: an eclipsing binary}\label{eclipse}
In Sect.\,\ref{err_pos} we mentioned that the Stripe\,82 data of SDSS\,J0106$-$0014 contains two extremely faint, but reliable detections.
Literature research revealed that SDSS\,J0106$-$0014 is a known eclipsing binary system \citep{kleinmanetal04-1}.
Previous observation of this objects reported a period of 0.085 days with a mid-eclipse time at MJD 55059.051
\citep{parsonsetal15-1}. This ephemeris  confirms that the dim detection of SDSS at MJD 53697.271 and  52522.362 indeed correspond to observations taken in eclipse.


\subsection{Magnetic White Dwarfs}\label{mwd_st82}
At least 4 per cent of all known white dwarfs have magnetic fields in the range 2$-$1000\,MG \citep{schmidt+smith95-1, liebertetal03-1, kepleretal13-1, Kleinmanetal13-1}. 
The mechanism that leads to the formation of strong magnetic fields in white dwarfs is still subject of debate, with several plausible mechanics being proposed: fossil fields conserved in the evolution of peculiar Ap and Bp stars (\citealt{angel+landstreet70-1}; \citealt{angel81-1}; \citealt{wickramasinghe+ferrario00-1}); fields generated by a magnetic dynamo during a phase of  binary evolution \citep{toutetal08-1, nordhouseetal11-1}; and fields generated in differentially rotating white dwarfs with convective envelopes \citep{markieletal94-1}.
  
Some magnetic white dwarfs also exhibit photometric modulation, most likely caused by stellar rotation combined with localized magnetic dichroism \citep{brinkworthetal13-1} or, in the case on convective white dwarfs (i.e $T_{\mathrm{eff}} \lesssim13,000$\,K for DAs), the  presence of star-spots \citep{lawrieetal13-1}.
The rotation period of magnetic white dwarfs can
potentially discriminate between different evolutionary scenarios and provide some insight on  the origin of the field.

\begin{figure}
\includegraphics[width=\columnwidth ]{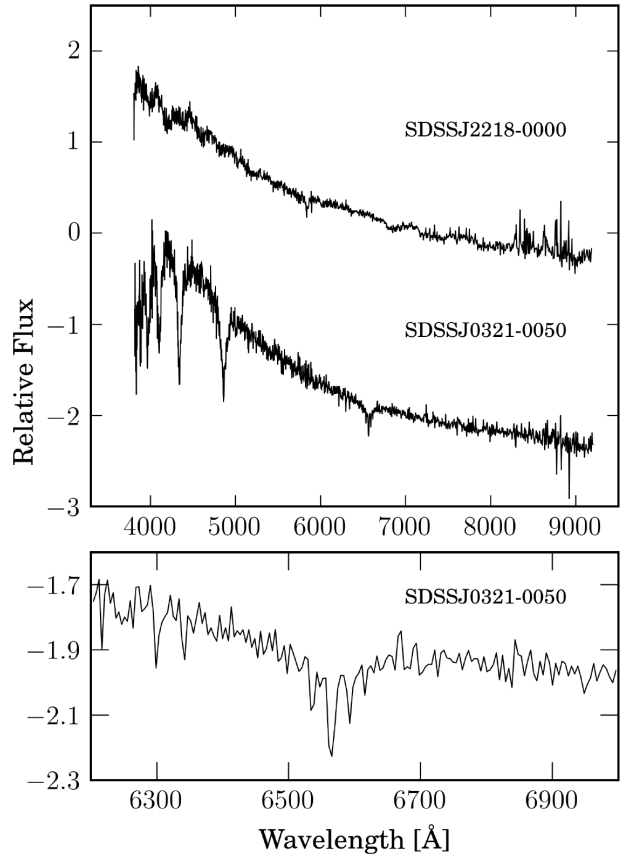}
\caption{\label{mag_spec} SDSS spectra of SDSS\,J2218$-$0000 and SDSS\,J0321$-$0050. In SDSS\,J2218$-$0000 the high magnetic field ($\simeq 225$\,MG) splits and significantly blurs the Balmer lines beyond recognition. In the case of SDSS\,J0321$-$0050 the presence of a weak field ($\simeq 1.4$\,MG) is revealed by Zeeman splitting of the H$\alpha$ absorption line (\emph{bottom panel}).}  
\end{figure}
Inspecting the 	SDSS spectra available for our 26 Stripe\, 82 variable candidates, we identify two magnetic white dwarfs: SDSS\,J2218$-$0000 and SDSS\,J0321$-$0050. The Zeeman splitting of H$\alpha$ in SDSS\,J0321$-$0050 is very weak (Fig.\,\ref{mag_spec}). In fact, this star was previously classified as a non-magnetic DA \citep{eisensteinetal06-1, Kleinmanetal13-1, gentilefusilloetal15-1}. Following \citet{reidetal01-1}, we estimate the average surface magnetic field strength, $B$s, according to the equation:
\begin{equation}
\hspace{2.5cm}
B\mathrm{s}/\mathrm{MG}=\frac{\Delta(1/\lambda)}{46.686},
\end{equation}
where $\Delta(1/\lambda)$ is the inverse wavelength separation in cm$^{-1}$ between the components of a Zeeman triplet \citep{reidetal01-1}; and find $B$s = $1.36\pm 0.04$\,MG.
 In contrast, SDSS\,J2218$-$0000 was already identified as a magnetic white dwarf and has a field sufficiently strong to smear out most of the Balmer lines ($B$s$\simeq225$\,MG, \citealt{schmidtetal03-1, kuelebietal09-1})




\subsection{The V777\,Her candidate SDSS\,J2333+0051}\label{other_pulse_st82}
Among our variable candidates we identified one DB white dwarf (	SDSS\,J2333+0051).
With only $\simeq20$ V777\,Her stars known to date, a robust definition of a DB white dwarfs instability strip, both empirical and theoretical, is still an ongoing challenge (\citealt{kilkennyetal09-1}, \citealt{nittaetal09-1}). Current evidence suggests that canonical 0.6\,M$_{\odot}$ DB white dwarfs undergo pulsation as they cool between roughly 29,000\,K and  21,000\,K \citep{nittaetal09-1}. SDSS\,J2333+0051 has $T_{\mathrm{eff}}=$  22,857\,K and $\log g =8.09$  (\citealt{Kleinmanetal13-1}; Table\,\ref{lit_par}) and is therefore a likely V777\,Her candidate.
As for our ZZ Ceti candidates,  we obtained high-speed photometry of SDSS\,J2333+0051 and carried out the required Fourier space analysis (Sect.\,\ref{lt_follow}).  We did not detect pulsations in SDSS\,J2333+0051 to a limit of 2.1 per cent. Based only on this result we cannot conclusively exclude that SDSS\,J2333+0051 is a V777\,Her, as it may have lower  amplitude pulsations. However the multi-epoch Stripe\,82 variability observed for SDSS\,J2333+0051 has a larger amplitude (0.16 $\pm$ 0.02 mag, Table\,\ref{phot_par}) than the limit obtained by our LT observations. It is therefore possible that this white dwarfs probably undergoes larger amplitude variations on longer timescales and we therefore encourage further observations of this star.

\subsection{The PG1159 star SDSS\,J0349$-$0059}
\label{PG1159}
By inspecting the available SDSS spectra we identified one of our variable candidate, SDSS\,J0349$-$0059, as a
PG1159 star (Table\,\ref{lit_par}). 
SDSS\,J0349$-$0059 is a well studied object and a known pulsator \citep{woudtetal10-1}. Being able to independently recover a pulsating PG1159 star further proves the reliability of our selection method.

\subsection{SDSS\,J2220$-$0041, a white dwarf plus brown dwarf binary}
\label{WD_BD}
One of our variable candidates, SDSS\,J2220$-$0041, is a known white dwarf plus brown dwarf binary (PHL 5038, \citealt{steeleetal09-1}). The presence of a low mass companion can, in some cases, cause some optical variability. Such variability is normally the result of irradiation of a tidally locked brown dwarf or ellipsoidal modulations (\citealt{littlefairetal14-1, casewelletal15-1, maozetal15-1}). Both these mechanisms require the stars in the binary to be very close, but SDSS\,J2220$-$0041 is instead a wide binary system. The two stars in the system are spatially resolved at an angular separation of 0.94"; corresponding to an orbital separation of $\simeq 55$ AU  \citep{steeleetal09-1}. 
The Stripe\,82 data alone does not allow to identify the nature of the variability or to speculate on a possible connection with the presence of the brown dwarf companion.

\subsection{The white dwarf candidate SDSS\,J2157+0037}
\label{phot_cand_st82}
Stripe\,82 has been the subject of many diverse studies and, consequently, it is one of the areas of the SDSS footprint with highest spectroscopic completeness. Indeed, of the 400 white dwarf candidates in our initial sample only 69 have no have a SDSS spectrum, and of the final 26 variable candidates found by our selection algorithm only one object, SDSS\,J2157+0037, lacks SDSS spectroscopy.
We can therefore only speculate about the nature of the observed variability. Its  $u-g (0.51)$ and  $g-r (-0.11)$ colours are compatible with those of ZZ Ceti stars (Fig.\,\ref{colour-comp}). Furthermore, the amplitude measured from its Stripe\,82 lightcurve is similar to those of the confirmed ZZ Cetis in our sample.
Again we encourage further observation of this object.

\section{Conclusions}
We have developed a  method to select variable white dwarfs in large-area time domain surveys.
Starting from a sample of white dwarf candidates, our method allows to correct and select reliable photometry using observations of neighbouring non-variable objects. Using this recalibrated photometry we then selected variable white dwarf candidates. We test our selection algorithm with a pilot search for pulsating  white dwarfs in the SDSS Stripe\,82.

From a sample of 400 white dwarf candidates taken from \citet{gentilefusilloetal15-1}, we identified 24 variable candidates.
From these 24 objects we further selected ZZ Ceti candidates using $u-g,g-r$ colours and acquired high-speed photometric follow up of six targets. We confirm five of our candidates as cool ZZ Cetis, three of which are new discoveries. Selection purely based on colour typically yields a success rate of only 13$-$30.  We show that non simultaneous multi-band photometry is one of the causes of this low efficiency as it leads to unreliable colours for cool pulsating white dwarfs.
However, we show that colour selection, combined with evidence of multi-epoch variability, significantly improves the quality of the ZZ Cetis candidates, without recourse to spectroscopy, achieving an efficiency of more than 80 per cent.

Among our candidates we also recover one known pulsating PG1159 star and one known eclipsing binary.
We speculate on the most likely cause for the observed variability of the remaining candidates. Even though we recommend further observations  to confirm beyond any doubt the variable nature of all our candidates, this pilot study already reveals the ability of our method to efficiently identify different types of variable white dwarf from eclipsing binaries to large amplitude cool pulsating white dwarfs.

SDSS Stripe\,82 proved a useful resource for testing our selection method, but covers only $300\,\deg^2$ of the sky, with sparse observations taken under variable observing conditions. Upcoming time-domain surveys will cover much larger areas of the sky with more continuous cadence (eg. PTF, \citealt{PTF09-1}; Pan-STARRS, \citealt{panstarss14-1}; Gaia, \citealt{Gaia_tran14-1}; LSST, \citealt{LSST11-1}). The application of our selection method to these surveys will lead to identification of variable white dwarfs on industrial scale and provide very large samples of high-confidence ZZ Ceti candidates. 
In this pilot study we were aided by the high spectroscopic completeness of Stripe\,82, but given the large area  coverage of future time-domain surveys, such intense spectroscopic follow-up may not always be available. The photometric selection method presented in \citet{gentilefusilloetal15-1} will perfectly complement future searches for variable white dwarfs providing large samples of high-confidence white dwarf candidates.

\section*{Acknowledgements}
NPGF acknowledges the support of Science and Technology
Facilities Council (STFC) studentships. The research leading to these results has received funding
from the European Research Council under the European
Union’s Seventh Framework Programme (FP/2007-2013) /
ERC Grant Agreement n. 320964 (WDTracer). 
We thank the referee for the constructive review.
Funding for SDSS-III has been provided by the Alfred
P. Sloan Foundation, the Participating Institutions, the
National Science Foundation, and the U.S. Department
of Energy Office of Science. The SDSS-III web site is
http://www.sdss3.org/.

\bibliographystyle{mn_new}

\end{document}